\documentclass[12pt]{article}
\usepackage{hyperref}
\usepackage{cite}
\usepackage{color}
\usepackage{graphicx}

\makeatletter
\@addtoreset{equation}{section}

\makeatletter
\renewcommand\section{\@startsection {section}{1}{\z@}%
                                   {-3.5ex \@plus -1ex \@minus -.2ex}%nn
                                   {2.3ex \@plus.2ex}%
                                   {\normalfont\large\bfseries}}
\renewcommand\subsection{\@startsection{subsection}{2}{\z@}%
                                     {-3.25ex\@plus -1ex \@minus -.2ex}%
                                     {1.5ex \@plus .2ex}%
                                     {\normalfont\bfseries}}

\def\baselinestretch{1.2}
\newcommand{\be}{\begin{equation}}
\newcommand{\ee}{\end{equation}}
\newcommand{\beq}{\begin{eqnarray}}
\newcommand{\eeq}{\end{eqnarray}}

\newcommand{\gone}[1]{{}}

\begin{document}
\begin{titlepage}
\begin{flushright}
MAD-TH-11-05
\end{flushright}
%\vspace{12 mm}

\vfil
%vfil

\begin{center}

{\bf \large 
Comments on domain walls in holographic duals of \\
mass deformed conformal field theories
}

\vfil

Akikazu Hashimoto

\vfil

Department of Physics, University of Wisconsin, Madison, WI
53706, USA

\vfil

\end{center}

%%%%%%%%%%%%%%%%%%%%%%%%%%%%%%%%%%%%%%%%%%%%%%%%%%%%%%%%%%%%%%%%%%%%%%%%%%%%%%%%%%%%%%%
\begin{abstract}
\noindent We consider M-theory backgrounds which are gravity duals of
mass deformed superconformal field theories in 2+1 dimensions.  The
specific examples we consider are the $B_8$, Stenzel, and the
Lin-Lunin-Maldacena geometries. These geometries contain compact
4-cycles on which one can wrap an M5-brane to create an object which
behaves effectively like a domain wall in 2+1 dimensions. We review
the quantization of flux and charges of these M-theory backgrounds, and
confirm that the back reaction of the domain wall shifts the charges
in a manner consistent with these quantization conditions, paying
particular attention to various subtle half integer shifts of the
charge lattice which arise as a part of the complete story. We also
describe a configuration of a stationary, merging M2/anti M2 pair in
the Lin-Lunin-Maldacena background, which can also be interpreted as a
domain wall, and compare its basic properties with the expectations from
its field theory description.
\end{abstract}
%%%%%%%%%%%%%%%%%%%%%%%%%%%%%%%%%%%%%%%%%%%%%%%%%%%%%%%%%%%%%%%%%%%%%%%%%%%%%%%%%%%%%%%%%
\vspace{0.5in}

\end{titlepage}
\renewcommand{\baselinestretch}{1.05}  %Line spacing
%%%%%%%%%%%%%%%%%%%%%%%%%%%%%%%%%%%%%%%%%%%%%%%%%%%%%%%%%%%%%%%%%%%%%%%%%%%%%%%%%%%%%%%%%%%%%

\section{Introduction}

Field theories in 2+1 dimensions are useful laboratories for exploring
dynamical issues in a framework that is well behaved in the
ultra-violet. Unlike in 3+1 dimensions, gauge field theories in 2+1
dimensions are super-renormalizable regardless of the number of
charged matter fields, and the duality cascades do not continue
indefinitely as one flows to the UV.

For some field theories, holography provides some additional tools to
probe their dynamical features. The prototype holographic duality for
field theories in 2+1 dimensions is the duality of ABJM
\cite{Aharony:2008ug} which relates $U(N)_k \times U(N)_{-k}$
Chern-Simons theory with bi-fundamental matter fields and a specific
superpotential to $AdS_4 \times S^7/Z_k$ geometry. This duality can be
understood as relating the decoupling limit of the world volume theory
on an M2 brane placed in the $R^8/Z_k$ transverse space to its gravity
description in the near horizon limit.

The ABJM duality has been generalized in a variety of ways including
changing the gauge group and the matter content, deforming the IR, and
scaling in new physics in the UV such as the Yang-Mills
interaction. In order to make the holographic duality precise, it is
important to correctly identify and relate the discrete and continuous
parameters appearing on both sides of the duality. Identification of
discrete parameters generally involve understanding quantized fluxes
and charges on the gravity side of the correspondence.

As we have seen in a number of examples, the task of quantifying and
discretizing fluxes and charges involves some subtleties. This stems
from the fact that there are several notions of charges, which in
simple contexts are indistinguishable, but can take on distinct values
and behave differently in more general settings. To avoid the
potential pitfall of confusing these subtle notions of charges, we
follow \cite{Marolf:2000cb} and use different names, brane, bulk, Page,
and Maxwell, in order to distinguish between them. These charges take
on different quantitative values in the presence of torsion and fluxes
when the space-time theory contain Chern-Simons terms, as is often the
case in supergravity theories. Page charges are integer quantized but
not gauge invariant. As such they map naturally to discrete parameters
such as level and rank which also exhibit ambiguity on the field
theory side via duality relations. Maxwell, brane, and bulk charges
are continuous and encode parameters and observables of the theory.  A
useful overview of the subtle roles of these charges can be found in
\cite{Aharony:2009fc}. One interesting outcome of the analysis of
charges for the ABJM theory is the prediction of a phase diagram in
charge space
\be N - {l (l-k) \over 2k} > 0 \ee
for the existence of a superconformal fixed point for the $U(N)_k
\times U(N+l)_{-k}$ theory.  It implies that the
Yang-Mills-Chern-Simons-Matter theory
\cite{Aharony:2009fc,Hashimoto:2008iv} (or some other UV embedding of
the Chern-Simons-Matter theory) with gauge group and level violating
this inequality must exhibit drastically different low energy
effective dynamics with spontaneously broken supersymmetry. The
precise nature of the low energy effective dynamics of this phase is
not currently well known, although there have been some 
attempts to investigate these issues
\cite{Hashimoto:2010bq,Bena:2010gs}.

In this article, we examine the issues which arise in gauge/gravity
correspondences when the superconformal field theories are mass
deformed in the IR. On the gravity side, these deformations generally
takes a background in a form of a cone such as the $R^8/Z_k$ and blows
up the point on the tip into a 4-cycle. We will focus on the class of
geometries whose supergravity solution are known explicitly: the $B_8$
geometry \cite{Gukov:2001hf}, the Stenzel geometry
\cite{Martelli:2009ga}, and the LLM geometry \cite{Lin:2004nb}. For
the LLM geometry, the dual field theory candidate is known very
explicitly \cite{Gomis:2008vc}. For the $B_8$ and Stenzel, the field
theory dual is not as well understood, but our analysis will not rely
on their details.  Our goal is to analyze the domain walls which arise
from considering M5-branes wrapping the blown up 4-cycle in these
geometries, which behaves effectively like a string of codimension one
from the point of view of the 2+1 extended dimensions.

The basic structure of such a domain wall was outlined in
\cite{Gukov:1999ya}. We will take a closer look at the vacuum on both
sides of the domain wall in specific setups listed above. There are
two main motivations for carrying out this exercise. One is to
diagnose the quantization conditions worked out for the fluxes and
charges in these backgrounds. If this was done consistently, the
charges and the fluxes should also be quantized accordingly on both
sides of the domain wall while preserving appropriate conserved
quantities. Since some of the ingredients for quantizing the fluxes
involved subtle issues
\cite{Aharony:2009fc,Hashimoto:2010bq,Gukov:2001hf,Hashimoto:2011aj}
such as Freed-Witten \cite{Freed:1999vc} and Pontryagin
\cite{Witten:1996md} anomalies, it would be a worth while exercise to
check the overall consistency of this scheme in the presence of the
domain wall. The other motivation stems from the close relation
between these domain walls and the tunneling effect which exists on
these blown up cones, originally identified in
\cite{Kachru:2002gs}. That a similar phenomenon exists on the Stenzel
background was shown recently in \cite{Klebanov:2010qs}, and it is not
difficult to see similar features also on the $B_8$ and the LLM
backgrounds. These transitions are particularly interesting in the
non-BPS context where a candidate meta-stable configuration appears in
the probe description of this system. It is interesting to clarify if
and when these metastable states are allowed to tunnel to a
supersymmetric vacuum.

This note is organized as follows. We begin by explaining the setup
and the subtleties for the case of $B_8$. Then, we will briefly
explain similar issues for the case of Stenzel. Finally, we will
examine the case of LLM. As a bonus, we identify a curious stationary
brane embedding in the LLM geometry which is interpretable as a probe
description of one of these domain walls.

\section{Domain wall in mass deformed $B_8$}

Let us begin our discussion by considering the case of warped mass
deformed $B_8$. The $B_8$ geometry, originally constructed in
\cite{Gibbons:1989er,Bryant:1989mv}, is an eight dimensional
non-compact manifold whose structure is that of a cone over squashed
$S^7$, deformed by blowing up the tip into an $S^4$. The resulting
geometry is Asymptotically Conical (AC).  The geometry can also be
viewed as a spinor bundle over $S^4$. This background supports a
normalizable anti-self-dual 4-form.\footnote{For the $B_8$ background,
we are adopting the convention of \cite{Aharony:2009fc} where the
anti-self-dual 4-form and an anti-M2 branes preserve the same
supersymmetry.}  When these structures are embedded as part of the
solution to eleven dimensional supergravity, the 4-form field strength
act as a source to back-react and warp the $B_8$ geometry. One can add
an anti-M2-brane localized in $B_8$ to further warp the geometry,
giving rise to an asymptotically $AdS_4 \times S^7_{squashed}$
space-time. It is common to take a $Z_k$ orbifold along the Hopf-fiber
of the squashed $S^7$ in the context of considering this geometry in
the context of AdS/CFT-like correspondences.

We should mention that there is another version of eight dimensional
space-time, also referred to as the $B_8$ geometry which asymptotes to
a circle fibered over a squashed $CP^3$ cone.  The squashed $CP^3$
arises naturally as the base of $U(1)$ fibration of the squashed
$S^7$. We will refer to these $B_8$ geometry as the Asymptotically
Locally Conical (ALC) $B_8$ geometries
\cite{Cvetic:2001pga,Cvetic:2001zx}.

The AC and ALC $B_8$ geometries are similar in that they are both
non-compact, admit normalizable anti-self-dual 4-forms, exhibit $spin(7)$
holonomy, and contain an $S^4$ Lagrangian 4-cycle. However, they
exhibit some differences in the asymptotic characterization of charges
and fluxes. This is not too surprising in light of the fact that these
spaces have different asymptotic geometries. We will therefore consider
the cases of AC and ALC asymptotics separately.

\subsection{Quantization of fluxes on  warped  $B_8^{AC}$}

In this subsection, we will describe the properties of domain wall
constructed by wrapping an M5-brane on the $S^4$ at the tip of
asymptotically conical $B_8/Z_k$ geometry. We begin by reviewing the
11 dimensional supergravity description of the warped $B_8/Z_k$
background in some detail.

We start with the Ricci-flat metric of the $B_8^{AC}$  
\be ds_8^2 = \left(1 - {\ell^{10/3} \over r^{10/3}}\right)^{-1} dr^2
+{9 \over 100} r^2 \left(1 - {\ell^{10/3} \over r^{10/3}} \right)
h_i^2 +{9 \over 20} r^2 d \Omega_4 \ , \label{ac} \ee
with
\be h_i \equiv \sigma_i - A^i_{(1)} \ , \ee
where $\sigma_i$ are left invariant one-forms on $SU(2)$, and
$A^i_{(1)}$ are $SU(2)$ Yang-Mills instanton on $S^4$. For $\ell = 0$,
this geometry reduces to a cone whose base is a squashed $S^7$
\cite{Duff:1983nu,Duff:1986hr}. The case with finite $\ell$
corresponds to deforming the tip of this cone so that there is a $S^4$
of finite radius at $r=\ell$.

This geometry admits an anti-self-dual 4-form field strength of the form
\be G_4 = d C_3 \ee
with 
\be C_3 = m \left( v_1(r) \sigma \wedge X_2 + v_2(r) \sigma \wedge Y_2
+ v_3(r) Y_3 \right) + \alpha d \sigma \wedge d \varphi \label{C3} \ee
and
\beq 
v_1(r) &=& {\ell^4  \over 5 r^4} + {4 \ell^{2/3} \over 5 r^{2/3}}\cr
v_2(r) & = & - {\ell^{2/3} \over r^{2/3}}  \\
v_3(r) &=& {\ell^{2/3} \over r^{2/3}}\ , \nonumber
\eeq
where $\sigma$, $X_2$, $Y_2$, $X_3$, and $Y_3$ are differential forms
on squashed $S^7$ defined in \cite{Cvetic:2001pga}.

One also can write the 4-form as
\beq dC &=& m \left[ u_1 (h a^2 b\, dr \wedge \sigma \wedge X_2 \pm
c^4 \, \Omega_4) + u_2 ( h b c^2 \, dr \wedge \sigma \wedge Y_2 \pm
a^2 c^2 \, X_2 \wedge Y_2) \right. \cr 
&& \left. +u_3 (h a c^2 \,dr \wedge Y_3 \mp a b c^2 \, \sigma \wedge
X_3)\right] \label{b18}\eeq
where
\be u_1 = -{800 \ell^{2/3} \over 27 r^{14/3}}, \qquad u_2 = -{400
\ell^{2/3} \over 81 r^{14/3}}, \qquad u_3 = {400 \ell^{2/3} \over 81
r^{14/3}} . \ee

The anti-self-dual 4-form sources negative M2 charge. This charge,
combined with the charges of additional anti-M2-brane added into the
system will give rise to a warp factor, which in the BPS ansatz will
take the form
\beq ds^2 &=& H^{-2/3} (-dt^2+dx_1^2 + dx_2^2) + H^{1/3} ds_8^2 \cr
F_4 & = & dt \wedge dx_1 \wedge dx_2 \wedge d \tilde H^{-1} + m
G_{4}\eeq
with $H$ solving the inhomogeneous condition
\be \nabla^2 H = {1 \over 2} G_4 \wedge G_4 + (2 \pi l_p)^6 Q_2^0
\delta^8(\vec r) \ee
although for simplicity, we will treat the delta function source to be
smeared along the $S^4$. The parameter $Q_2^0$ is the brane charge
which we will relate to with the various discrete parameters
below. Let us point out for now that the brane charge $Q_2^0$ is not
the same as the Page charge $N$.

Let us now describe the quantization of parameters in this
supergravity background.

First, we quantize the magnitude of the anti-self-dual 4-form as
follows. Because $S^4$ at $r=\ell$ is a closed surface in this
geometry when $\ell > 0$, one must impose the quantization of the
period of $G_4$ pulled back on $S^4$.  This condition reads
\be \left.\int_{S^4} m u_1 c_4 \Omega_4 \right|_{r=\ell} = - 16 \pi^2
m = (2 \pi l_p)^3 \left(q-{k \over 2}\right) \ee
where $q$ is an integer. Here, the shift by $k/2$ is due to the effect
of \cite{Witten:1996md} whose presence in the context of the $B_8$
geometry was argued in \cite{Gukov:2001hf}.

Somewhat less obvious is the quantization of $\alpha$ in (\ref{C3})
which characterizes the torsion class of the 3-form potential
restricted to the squashed $S^7/Z_k$ boundary of the $B_8^{AC}$.  This
is most apparent if one sets $\ell=0$ and $m=0$ so that we are left
with a cone over squashed $S^7/Z_k$. Just as was the case for the
example in \cite{Hashimoto:2011aj}, it is somewhat subtle to read off
the discrete torsion when $\ell$ and $m$ are non-vanishing. One quick
way to read off this quantity is to take $r\rightarrow \infty$ where
$G_4$ goes to zero. This imposes the constraint
\be 16 \pi^2 \alpha = -(2 \pi l_s)^3 g_s \left(l - {k \over 2} \right)
\ee
where the shift by $k/2$ due to the Freed-Witten anomaly has been
included since the IIA reduction of squashed $S^7$ is a squashed
$CP^3$ which contains a non-spin $CP^2$ homology 4-cycle.

Note, however, that the period of $G_4$ on any 4-cycles on $S^7/Z_k$
or its IIA reduction vanishes since the angular component of $G_4$
decays sufficiently rapidly for large $r$. In other words, the D4
Maxwell charge in the IIA reduction vanishes.

We next consider the quantization of M2 charges. It is useful to first
consider the case where we set $N=0$.  The background will nonetheless
carry non-vanishing M2 charge because of the fluxes and the discrete
torsion.

The equation for $G_4$ reads
\be d *_{11} G_4 = {1 \over 2} G_4 \wedge G_4 \label{eqmG4} \ . \ee
Let us denote the flux of $*G_4$ at $r=\infty$ and $r=\ell$ as
\beq Q_2^{\infty} &=& \left. {1 \over (2 \pi l_p)^6} \int *G_4
\right|_{r = \infty} \\ 
Q_2^{0} &=& \left. {1 \over (2 \pi l_p)^6}\int *G_4 \right|_{r=\ell} \
.  \eeq
The equation of motion (\ref{eqmG4}) constrains their difference 
\beq Q_2^{\infty} - Q_2^0 &=& {1 \over (2 \pi l_p)^6} \int_{{\cal
M}_8} {1 \over 2} G_4 \wedge G_4 \cr
& = & - {\left(q-{k \over 2}\right)^2 \over 2k } \eeq
which must be negative in order not to break additional
supersymmetries because $G_4$ is anti-self-dual. Since $q$ is an
integer and this quantity can never be set to zero, there is always
some non-vanishing M2 charge associated with this background even
though we set $N=0$.

Now, in the presence of discrete torsion, one expects 
\be Q_2^\infty = -{(l - {k \over 2})^2 \over 2k} + {k \over 8} =
-{l(l-k) \over 2k}\ee
where the additive term $k \over 8$ was included to arrange for this
contribution to vanish for $l=0$. With this parametrization, the
domain wall\footnote{This domain wall is localized in the radial
direction of the gravity solution and should not be confused with the
domain wall on the 2+1 dimensional field theory which is the main
subject of this paper.}  described in section 3.4.1 of
\cite{Aharony:2009fc} will have the correct induced charge.

This means
\be Q_2^0 = -{l(l-k) \over 2k} + {\left(q-{k \over 2}\right)^2 \over
2k } \ee
which can also be written as
\be Q_2^0 = {k \over 8} + \left(l - {k\over 2} \right) b_0 + {k \over
2} b_0^2 \ee
where 
\be b_0 = -{l-q \over k} \ee
is the value of $B$ pulled back on to $S^2$ in the IIA reduction of
$R^4/Z_k \rightarrow R_+ \times S^2$.

If one were to add $N$ units of M2-brane charge at $r=\ell$ (and
smeared along the $S^4$) one finds
\be Q_2^0 = N - {l(l-k) \over 2k} + {\left(q-{k \over 2}\right)^2
\over 2k } \label{Q0} \ee
and
\be Q_2^\infty = N - {l(l-k) \over 2k} \label{Qinf} \ee
where negative $N$ corresponds to the branes which preserves the same
supersymmetry as the anti-self-dual 4-form field strength $G_4$. The
condition that these backgrounds are BPS can be expressed simply as
\be Q_2^0 < 0 \ . \ee
Naive extrapolation of the $Q_2^0<0 $ solutions to the $Q_2^0>0$
region introduces a repulson singularity indicating that the basic
ansatz used to construct the solution is breaking down
\cite{Aharony:2009fc}. It is in light of this fact that it is
interesting to consider the fate of a probe M2 added to the background
with $Q_2^0=0$ which pushes the system just beyond the threshold of
supersymmetry breaking. We will come back this issue later in this
article.

Since the language and the notation being used here is somewhat
different from that of \cite{Gukov:1999ya}, let us provide the map
\be Q_2^\infty = \Phi^{GVW}, \qquad Q_2^0 = N^{GVW} \ee
for the quantity $\Phi^{GVW}$ and $N^{GVW}$ introduced in (2.16) of
\cite{Gukov:1999ya}. In the language of \cite{Marolf:2000cb},
$Q_2^\infty = \Phi^{GVW}$ is the Maxwell charge which is required to
remain constant as one crosses the domain wall.  For $k>1$, the
quantity $Q_2^0=N^{GVW}$ is taking on fractional values. These
correspond to the brane charge in the language of \cite{Marolf:2000cb}
and are allowed to take on fractional values.  For $k=1$, $Q_2^0$
actually takes on integer value up to an additive shift by $k/8$
which, if desired, can be absorbed into the shift of the charge
lattice.

\subsection{Domain wall from wrapped M5 in $B_8^{AC}$}

Let us now consider what happens when one wraps an M5-brane on the
$S^4$ Lagrangian cycle at the tip of the deformed $B_8^{AC}/Z_k$
cone. The $B_8^{AC}/Z_k$ geometry has the structure of $R^4/Z_k$
fibered over $S^4$, and wrapping an M5-brane on the $S^4$ will give
rise to an object which looks effectively as a codimension one string
along the 2+1 extended dimensions.

The issue which one must address is the basic phenomenon that the
$G_4$ flux on the world volume of M5 induces anomalous world volume
charge which must be canceled by the correct number of open M2-branes
ending on the M5.  This is the same basic mechanism which gave rise to
open strings ending on the baryon vertex in the construction of
\cite{Witten:1998xy}. The precise number of open M2-branes required to
cancel this anomaly is the quantized period of $G_4$ on $S^4$.  This,
however, is somewhat problematic since we saw that on $B_8$, the flux
of $G_4$ on $S^4$ can take half integral values when $k$ is odd. On
the first glance, it would appear that half integer unit of open M2 is
required to properly cancel the anomaly when $k$ is odd.

With a little more thought, however, one realizes that this potential
problem is already addressed in \cite{Gukov:1999ya}. The resolution is
as follows. If an M5 is wrapped on the $S^4$, it must also give rise
to a shift in the flux of $G_4$ by one unit in its dual cycle, which
in this case is $R^4/Z_k$. Then, on the covering space $R^4$, the
shift of flux is $k$. On the other hand, because the BPS vacuum
configuration of $G_4$ must be anti-self-dual on the covering space,
$q$ must shift by $k$ unit as one crosses the domain wall.

If the flux of $G_4$ on $S^4$ is $q-k/2$ on one side of the domain
wall and $q+k/2$ on the other side, then their average value is
$q$. To refer to the average value of the flux seems quite natural
from the point of view of taking the thin wall approximation of the
domain wall, and is what is employed in \cite{Gukov:1999ya}.  So $q$
is the number of anomalous open M2 one expects to need in order to
cancel the anomalous charge on the world volume of M5 wrapping the
$S^4$, and this quantity is integer valued.

One can characterize the vacua separated by the domain wall as
follows: Suppose on one side of the domain wall, we were provided with
the data $N$, $l$, $k$, and $q$ so that $Q_2^0$ and $Q_2^\infty$ are
as is given in (\ref{Q0}) and (\ref{Qinf}), respectively. Upon
crossing the domain wall, $q$ shifts to $q+k$. In a mean time, the
brane charge
\be (Q_2^0)' = N - {l(l-k) \over 2k} + {(q+{k\over 2})^2 \over 2k} =
Q_2^0+q \ee
shifts by $q$ units.  The Maxwell charge $Q_2^\infty$ is invariant
under wall crossing as is expected.  The physical picture arising from
interpreting the structure of the vacuum on both sides of the domain
wall appears to be consistent with the charge quantization conditions
outlined in the previous section.

\subsection{Domain wall from wrapped M5 in $B_8^{ALC}$}

In this subsection, we will describe the features of a domain wall
arising from wrapping an M5-brane on the blown up $S^4$ at the tip of
asymptotically locally conical $B_8$. The $B_8^{ALC}$ has two physical
scales, one associated with the radius of the blown up $S^4$, and the
one associated with the radius of $S^1$ at infinity. For a particular
numerical value for the dimensionless ratio of these two scales, the
$B_8^{ALC}$ geometry can be viewed as an analytic continuation of
another asymptotically locally conical geometry, known as the $A_8$
geometry, and has a relatively simple analytic form
\cite{Cvetic:2001pga}. Generalizing to the case where the ratio of the
radius of $S^4$ and $S^1$ leads to a somewhat more implicit form of
the metric and the flux, but they are known
\cite{Cvetic:2001pga,Cvetic:2001zx}. We will refer to this one
parameter family of $B_8^{ALC}$ geometry as $B_8^{ALC}(\lambda)$ where
$\lambda$ parametrizes the ratio of the radii as was reviewed in
appendix B of \cite{Hashimoto:2010bq}.

The quantization of fluxes and the interpretation of various distinct
notion of charges for the $B_8^{ALC}(\lambda)/Z_k$ was described in
\cite{Hashimoto:2010bq}.  It was shown in (5.23) of
\cite{Hashimoto:2010bq} that the M2 Maxwell charge takes the form
\be Q_2^{\infty} = N - {l(l-k) \over 2k} + {1\over 2k} \left({4
C(\lambda) \over 25}\right) \left(q-{k \over 2}\right)^2
\label{B8lamQ2inf} \ee
where $C(\lambda)$, partially illustrated in figure 11.a and figure
12 of \cite{Hashimoto:2010bq}, takes value ranging from zero to
infinity as the ratio of radii of $S^4$ and $S^1$ are varied.

Now, imagine wrapping an M5-brane on $S^4$ in the core region of
$B_8^{ALC}(\lambda)/Z_k$. Topologically, $B_8^{ALC}(\lambda)/Z_k$ has
the structure, identical to $B_8^{AC}$, of an $R^4/Z_k$ fibered over
the $S^4$ base. As such, just as was the case for the $B_8^{AC}$, one
expects the insertion of domain wall in the 2+1 extended dimensions by
wrapping an M5 brane on the $S^4$ to cause the flux of four form
through $R^4/Z_k$ and $S^4$ to jump by one unit across the domain
wall. In other words, the value of $q$ should jump by one across the
domain wall.

Now, since the M2 Maxwell charge (\ref{B8lamQ2inf}) appears to depend
explicitly on $q$, one may wonder if this jump is causing the M2
Maxwell charge to also jump across the domain wall.  This would be
contrary to the expectation based on the fact that Maxwell charge is
supposed to be a conserved quantity which maps to a parameter defining
the theory, as opposed to specifying the vacuum, in the corresponding
field theory. This point has caused some confusion, especially to the
present author, during the earlier stages of studying this issue.

It turns out that the resolution to this confusion is quite straight
forward.  There is no need to assume that the value of $\lambda$ are
the same on two sides of the domain wall. In fact, as the flux of
$G_4$ though the $S^4$ changes, one expects the radius of $S^4$ to
change relative to the radius of $S^1$. Since radius of $S^1$ is fixed
at ultra-violet, changes in the radius of $S^4$ will cause $\lambda$
to change. Since the M2 Maxwell charge (\ref{B8lamQ2inf}) must be
conserved, we expect $\lambda$ to adjust itself accordingly.  In the
case of $B_8^{AC}$, this issue did not arise simply because there is
no scale relative to which one can measure the radius of the $S^4$.

\section{Domain wall in Stenzel Geometry}

As our second example, let us consider the properties of domain walls
in Stenzel geometry \cite{Stenzel}. Stenzel geometry is a deformation
of the cone with the base $V_{5,2}$ known as the Steifel manifold, by
blowing up an $S^4$.  The embedding of the warped deformed Stenzel
geometry into string theory was first considered in
\cite{Cvetic:2000db} and further elaborated in the context of AdS/CFT
correspondence in \cite{Martelli:2009ga}. This geometry can also be
viewed as a contangent bundle over $S^4$. Asymptotically, Stenzel
geometry has the structure of a cone. One difference between this case
and the $B_8^{AC}$ is that the $Z_k$ orbifold along the $U(1)_b$
isometry of the asymptotic $V_{5,2}$ does not act freely on the $S^4$.
Nonetheless, one can consider wrapping an M5-brane on the $S^4$ in
this background and interpret it as a domain wall, as was done in
\cite{Klebanov:2010qs}.  This construction was also treated as the
prototype in \cite{Gukov:1999ya}.

The quantization and computation of various charges for the Stenzel
geometry was carried out in \cite{Hashimoto:2011aj}. We follow the
convention of \cite{Hashimoto:2011aj} in setting up the geometry and
the flux. The M2 brane and Maxwell charge was found in (3.16) and
(3.18) of \cite{Hashimoto:2011aj} to take the form
\be Q_2^0 = N - {l (l - 2k) \over 4k} - {k q^2 \over 4}
\label{stzQ2brane}\ee
and 
\be Q_2^\infty = N-{l(l-2k) \over 4k} \ . \ee
The parameter $q$ is the integer quantized period of the self-dual
4-form through the $S^4$.  There are no anomalous shift by $k/2$ as
was the case for the $B_8$ in the previous section. Since the four
from is taken to be self-dual, the condition for supersymmetry is for
$Q_2^0$ to be positive.

The issue we wish to address now is whether these expressions are
compatible with the expected properties of the embedding of the
M5-branes.

The first issue we need to establish is the extent to which we expect
the $q$ to shift across the domain wall. For concreteness, let us
concentrate on the case where $k=1$ so that we can ignore the effects
of orbifolding by $Z_k$. Then, we do not expect fractional brane
sources to exist in the background.

Now consider shifting $q$ by one in (\ref{stzQ2brane}). This gives
rise to
\be \Delta Q_2^0 = -{(q+1)^2 \over 4} + {q^2 \over 4} = -\left({q
\over 2} + {1 \over 4}\right) \ee
which is not integral. Note, however, that if one shifts $q$ by 2,
then the shift in (\ref{stzQ2brane}) 
\be \Delta Q_2^0 = -{(q+2)^2 \over 4} + {q^2 \over 4} =  -(q+1)
\ee
is integral. This turns out to make sense. For one unit self-dual
4-form $G_4$ normalized so that
\be \int_{S^4} G_4 = 1 \ee
we have
\be \int_{R^4} G_4 = {1 \over 2} \ee
as can be seen in (3.9) and (3.10) of \cite{Hashimoto:2011aj} and
originally computed in (5.79) and (5.85) of
\cite{Martelli:2009ga}. This means that if crossing the domain wall
made by wrapping an M5 on $S^4$ were to cause the flux of $G_4$ on the
dual $R^4$ cycle to jump by one unit, then the flux of $G_4$ though
$S^4$ must also jump by two units.  Shifting $q$ by two is precisely
what we found as giving rise to an integral shift in
(\ref{stzQ2brane}). Furthermore, if the value of $q$ is shifting by 2
units, then taking the average of flux though $S^4$ on both sides of
the domain wall gives
\be {1 \over 2} \left( (q+2) + q \right) = q+1 \ee
which is precisely the shift found in (\ref{stzQ2brane}) up to a sign
which sets the orientation of the M5-brane.  All of the numerical
details including the factors of 2 conspire to yield a self-consistent
picture.

In the case of the $Z_k$ orbifold, one can work in the covering space
for which the shift of the brane charge now reads
\be k \times  \Delta Q_2^0 = -k^2 (q+1) \ . \ee
If before crossing the domain wall the flux of $G_4$ on $S^4$ in the
covering space is $k q$, and on the other side the same flux had
jumped to $k(q+2)$, then their average is $k(q+1)$, and insertion of
$k$ M5-brane will induce precisely $k^2 (q+1)$ units of open M2 charge
from the world volume anomaly mechanism, once again giving rise to a
consistent picture.

We are therefore finding that the quantization of fluxes and charges
which were prescribed in \cite{Hashimoto:2011aj} is consistent with
the interpretation of M5-branes wrapped on $S^4$ as domain walls
interpolating between consistent supersymmetric vacua in the
holographic description of these 2+1 dimensional field theories.

\section{Domain wall in Lin Lunin Maldacena background}

As a final example, we will consider the properties of domain walls
arising from M5-branes wrapping 4-cycles in the asymptotically $AdS_4
\times S^7$ geometry of Lin, Lunin, and Maldacena
\cite{Lin:2004nb}. There are few features which makes this example
different from the examples considered earlier in this article. The
most important difference perhaps is the fact that the Lagrangian of
the conjectured field theory dual of this geometry is known very
explicitly \cite{Gomis:2008vc}, providing opportunities to carry out
many detailed comparisons.

\subsection{Review of the LLM background}

Let us begin by reviewing some of the basic features of the LLM
solution, which was originally developed to characterize the solutions
of type IIB supergravity which were asymptotically $AdS_5 \times
S^5$. They were able to classify these solutions in terms of a diagram
consisting of a flat two dimensional plane, colored in patches by one
of two colors, say black and white. In the case where the region
covered by one of the colors have finite area, the diagram corresponds
to a specific type IIB geometry asymptoting to $AdS_5 \times S^5$
whose radius is proportional to the area of the patch in the two
dimensional plane.

\begin{figure}[t]
\centerline{\includegraphics{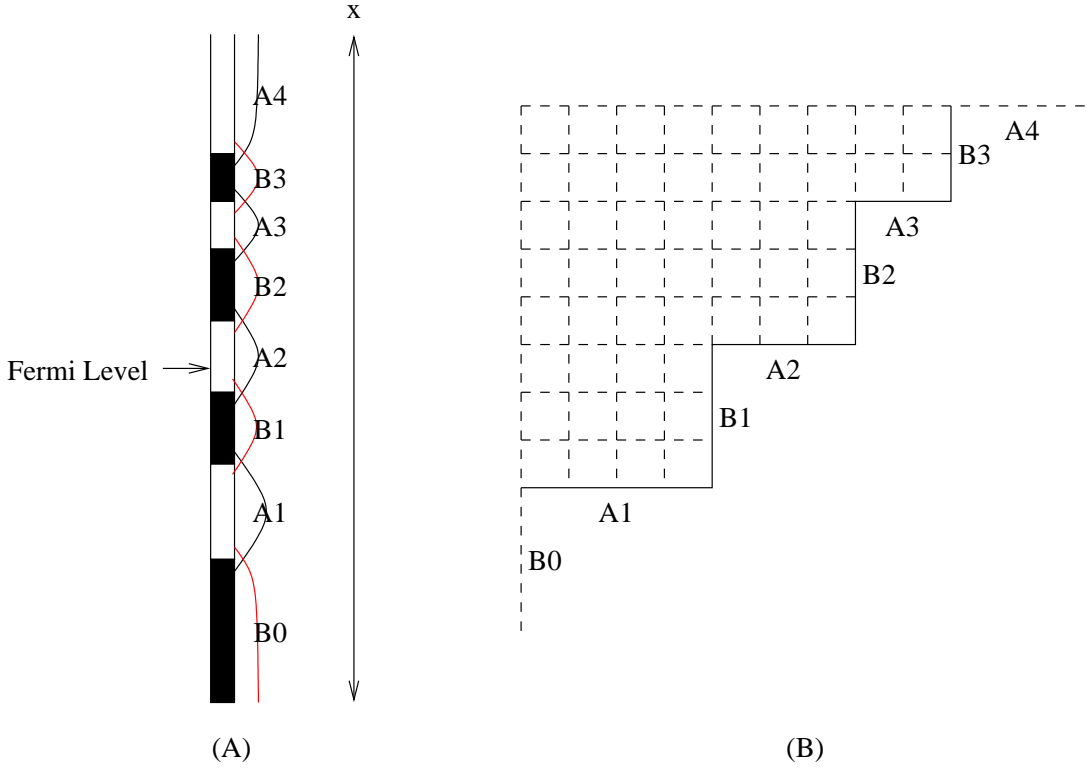}}
\caption{(A) A typical bubble diagram for the asymptotically $AdS_4
\times S^7$ LLM geometry, and (B) the corresponding Young diagram.
\label{figa}}
\end{figure}

One can also construct a family of asymptotically pp-wave geometries
by taking the Penrose limit of the family of asymptotically $AdS_5
\times S^5$ geometries described above.  These solutions are
characterized by the same two dimensional diagram, but with the
colored regions exhibiting translation symmetry in one of the
coordinates, so that the colored region look somewhat like a bar-code.
The transitionally invariant direction along the bar-code diagram
corresponds to an isometric direction in the IIB geometry it is
representing. One can therefore construct a family of solutions to
11-dimensional supergravity by compactifying that dimension,
T-dualizing along this coordinate, and lifting to M-theory.  These
solutions are characterized by a one dimensional strip like the one
illustrated in figure \ref{figa}.A. It is customary to give the strip
a finite width to make it easier to read. If the strip is such that it
asymptotes to a solid black region on one end and a solid white region
on the other, the 11 dimensional supergravity solution will be
asymptotically $AdS_4 \times S^7$. To avoid complicating the
discussion, we will restrict to the case where we do not perform any
$Z_k$ orbifolding, i.e.\ set $k=1$. A nice discussion of these
structures in the case of $k>1$ can be found in \cite{Cheon:2011gv}.

The direction along the strip parametrized by coordinate $x$ is
embedded as one of the coordinates of the M-theory background. In
addition, there are semi-infinite coordinates $y$ and the coordinates
of two 3-spheres $S^3$, $\tilde S^3$ parametrizing the 8 dimensions
transverse to $R^{1,2}$, whose detailed form can be found in
(2.33)--(2.35) of \cite{Lin:2004nb}. In this construction, the black
and white regions correspond to the points in $x$ where $S^3$ or
$\tilde S^3$, respectively, shrinks to zero size as $y \rightarrow 0$.

This implies that there is an abundance of 4-cycles in this
geometry. Any segment straddling a black or white region embedded in
the $xy$ plane whose endpoints are constrained to have $y=0$ will
define a 4-sphere of the form $S^4 = I \times S^3$ or $S^4=I \times
\tilde S^3$. So these 4-cycles are in one to one correspondence with
the finite size strips labeled A1, B1, ... B3 in figure \ref{figa}.A.
These 4-cycles link, as is also illustrated in figure
\ref{figa}.A. The semi-infinite black and white strips labeled as B0
and A4 in figure \ref{figa}.A defines a 4-cycle which is non-compact.

The flux of M-theory 4-form though these 4-cycles are quantized.  This
turns out to force the length of the colored strip to be quantized as
well. It was found in (2.24) of \cite{Cheon:2011gv} that the length of
the finite size strips are integer multiples of
\be {(2 \pi l_p)^3 \mu_0 \over (2 \pi)^2} \label{quant} \ee
where $\mu_0$ is a scale parameter included explicitly in the
presentation of LLM solution in (2.11) of \cite{Cheon:2011gv} which we
find to be convenient for comparing supergravity and field theory
computations.

As long as the two semi-infinite strips, i.e.\ B0 and A4 in figure
\ref{figa}.A is of different color, one can define the Fermi-level
which is the point along the strip where the areas, of the black
region above and of the white region below this point, are the
same. If the black and white regions are interpreted as denoting
occupied and unoccupied states in a degenerate fermi-gas system, this
point would in fact be interpretable as the Fermi-level of this
system.  One can also consider the total energy carried by the fermion
and hole excitations above the ground state of this degenerate
fermi-gas. This quantity turns out to correspond to the radius of the
asymptotic $AdS_4 \times S^7$ region.  In other words, it is encoding
the Maxwell charge. Transition between different vacua of the same UV
field theory corresponds to transition between different excited
fermion occupation states with the same total energy.

Another useful way to represent the black and white strip is in the
form of a Young tableaux where the white and black strips are mapped
to horizontal and vertical edges of the tableaux, respectively, as is
illustrated in figure \ref{figa}.B. In this diagram, the number of
boxes correspond to the total energy of the fermi gas and therefore
the Maxwell charge of the supergravity background. Young tableaux
corresponding to different vacua of the same theory will therefore
have the same number of boxes.

The field theory counterpart of these vacua is also well
understood. The candidate dual field theory is the mass deformed ABJM
theory constructed by \cite{Gomis:2008vc,Hosomichi:2008jb}. In order
to be completely specific about the conventions including the
normalization of the mass parameter $\mu$, we will consistently use
the model written down in (2.1) and (2.2)
\cite{Gomis:2008vc}.\footnote{We warn the reader that appendix C.1 of
\cite{Gomis:2008vc} appears to deviate from their own conventions by
various factors of $\pi/k$. Appendix C.2, on the other hand, appears
to be consistent with the rest of their article. We suspect the
unexpected factors of $\pi/k$ which can be seen, for example, by
comparing their expressions for $V$ at the end of section C.1 and C.2,
is a typo.}  It was pointed out by various authors including
\cite{Gomis:2008vc,Cheon:2011gv} that the classical vacua of the mass
deformed ABJM model can be expressed in a block-diagonalized form
illustrated in figure 1 of \cite{Cheon:2011gv}, where each of the
sub-blocks are either of dimension $n \times (n+1)$, or of $(n+1)
\times n$.  The $n$'s are allowed to take on any non-negative integer
values, and it is not difficult to derive a relation from the fact
that these sub-blocks should fit inside an $N\times N$ matrices that
\be \sum_{n=0}^\infty \left(n +{1\over 2}\right) (N_n + N'_n) = N,
\qquad \sum_{n=0}^\infty N_n = \sum_{n=0}^\infty N'_n \ . \ee
In other words, $N_n$ and $N'_n$ are naturally interpretable as
particle occupation numbers. They will correspond perfectly with the
fermion occupation numbers illustrated in figure \ref{figa}.A if the
values of $N_n$'s and $N'_n$'s are constrained to only take values 0
or 1. (This statement has a simple generalization for the case $k>1$
\cite{Cheon:2011gv}.) From the field theory point of view, however,
$N_n$'s and $N'_n$'s can take arbitrary integer values and still give
rise to a solution to the equation of motion.  This mismatch is one of
the subtle unresolved issues in the holographic duality of this
system. The general consensus is that the classical vacua with $N_n$'s
and $N'_n$'s taking values larger than one do not exist as a state in
the quantum theory. Computation of the Witten index appears to confirm
this picture \cite{Kim:2010mr}. At the moment, we are missing the
understanding of the precise mechanism which destabilizes or lifts
these classical vacua which allegedly do not exist at the quantum
level. In the remainder of this article, we will adopt the point of
view that these states with $N_n$'s and $N'_n$'s taking values greater
than one are indeed absent in the quantum theory.

\subsection{Domain wall from wrapped M5-branes}

\begin{figure}[t]
\centerline{\includegraphics{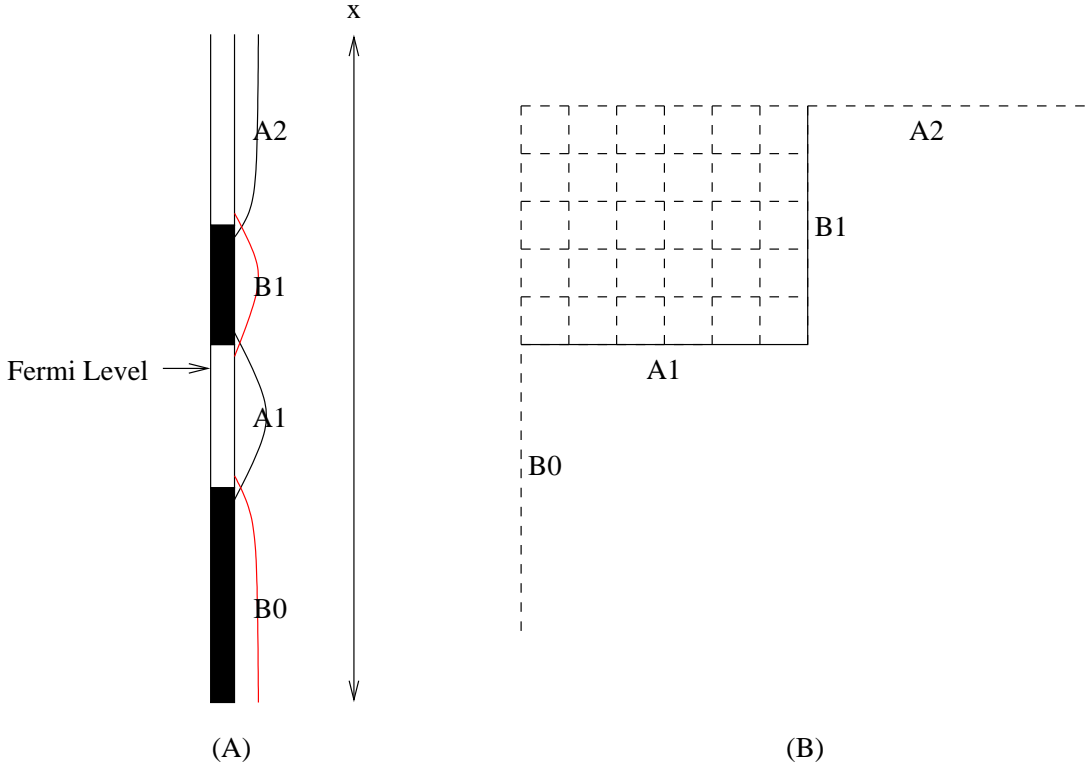}}
\caption{A simple configuration of the asymptotically $AdS_4 \times
S^7$ LLM bubble geometry with compact 4-cycles A1 and B1, and the
corresponding Young diagram. \label{figb}}
\end{figure}

Let us now discuss the properties of domain walls constructed by
wrapping an M5 branes in one of the many 4-cycles which exists in the
LLM background.  To be more concrete, let us consider a specific
simple configuration of the bubble diagram illustrated in figure
\ref{figb}.A. This is the minimal setup for our purposes. A seemingly
smaller setup consisting only of black and white semi-infinite strips
with no islands have vanishing $N$ and is a singular
geometry.\footnote{One could also  consider a configuration from
appendix D of \cite{Lin:2004nb} but that would destroy the $AdS_4
\times S^7$ asymptotics and force us to scale $N$ to infinity.}

\begin{figure}[t]
\centerline{\includegraphics{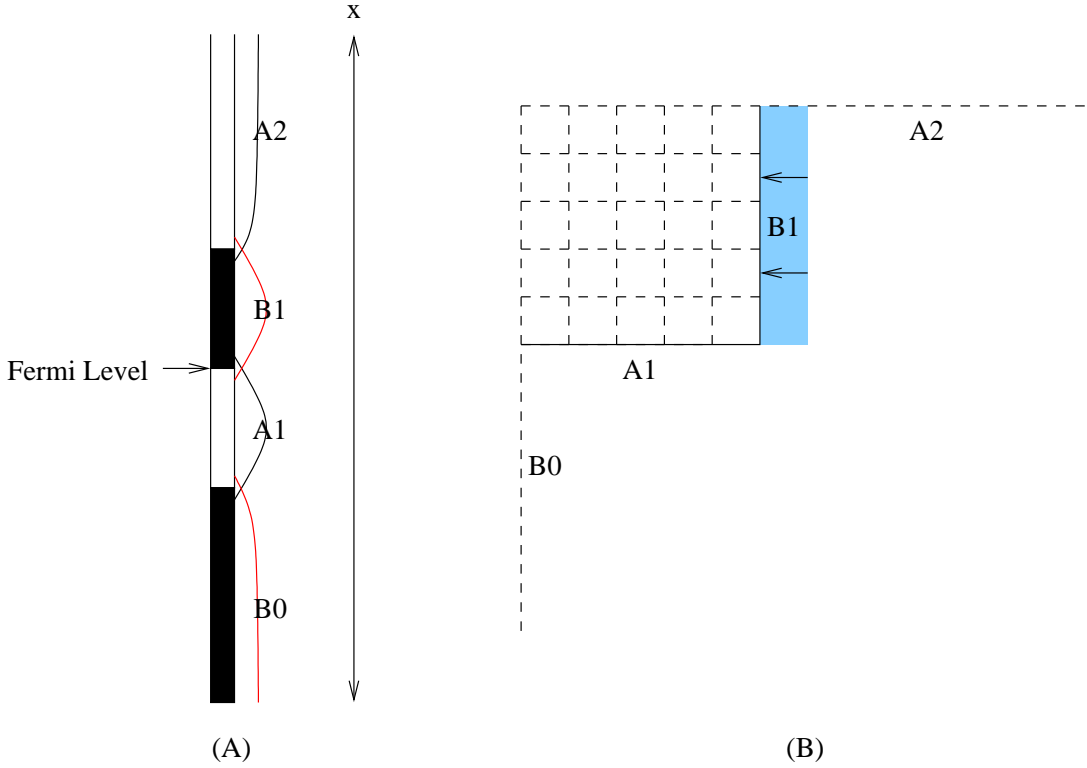}}
\caption{Effects of the back reaction of wrapping an M5-brane on
B1. \label{figc}}
\end{figure}

Let us now imagine wrapping an M5-brane around the 4-cycle labeled as
B1 in figure \ref{figb}. The cycles dual to B1 which links to it are
easy to identify. They are the cycles A1 and A2.  If as a result of
the M5 wrapping B1 the flux though A1 and A2 decrease, and increase,
by one, respectively, we will arrive at a new configuration
illustrated in figure \ref{figc}.

Unlike in the cases of $B_8$ and Stenzel manifolds, the background
4-form flux is not self-dual. Shifting the flux though A1 and A2 as we
here do not force the flux of B1 to also shift. The configuration
illustrated in figure \ref{figc} corresponds to a perfectly sensible
background.

One issue with the configuration of \ref{figc}, however, is the fact
that the number of boxes is not the same as the one from figure
\ref{figb}.  This is because we have yet to account for the open M2
branes which need to terminate on the M5 to cancel the anomalous world
volume charge. That number, $n_{B1}$, corresponding to the flux though
B1, is precisely the number of boxes deleted in figure \ref{figc}.

\begin{figure}[t]
\centerline{\includegraphics{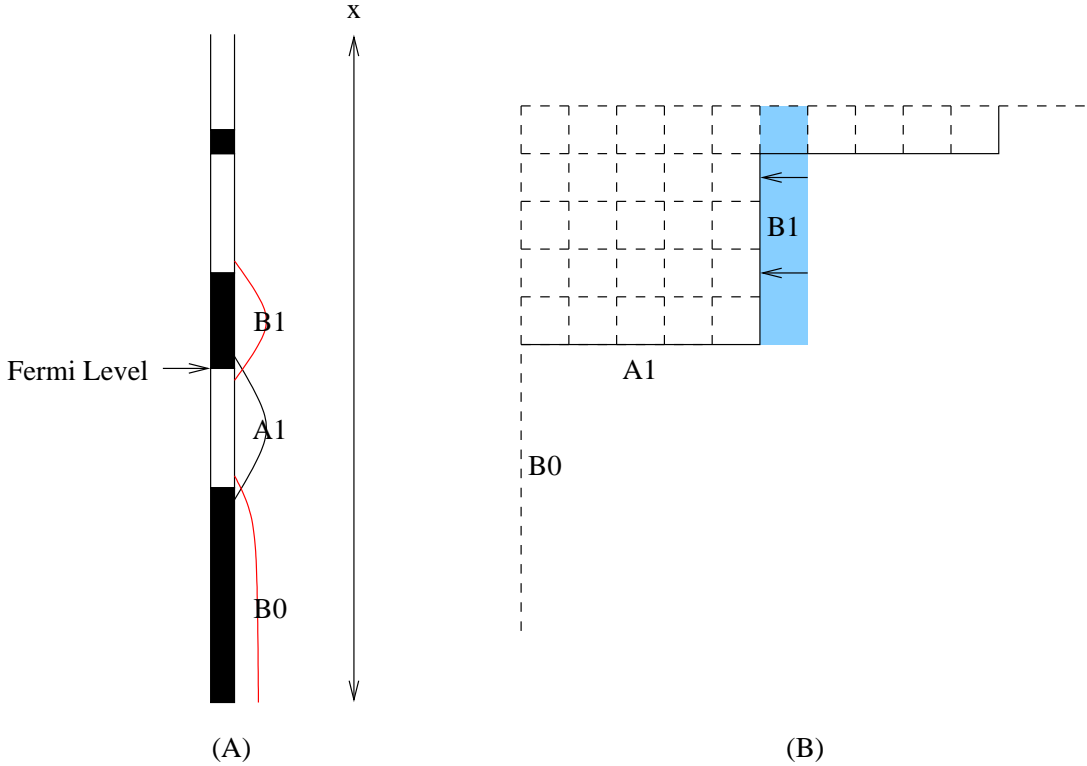}}
\caption{Effects of back reaction of wrapping the M5 on B1 and the
$n_{B1}$ open M2 needed to cancel the induced charges on M5 along B1.
There are many possible ways to arrange the M2. This particular
configuration allows the M2's, which combine to form a dielectric M5,
to be treated in the brane probe approximation. \label{figd}}
\end{figure}

There are variety of ways in which the $n_{B1}$ boxes can be added
back to the configuration of figure \ref{figc}. One example is
illustrated in figure \ref{figd}. The point of this configuration is
that in the limit where the number of rows and column represented by
the flux through A1 and B1 are large, the $n_{B1}$ additional boxes
can be realized in the gravity description effectively in a probe
description. Generally, long strips have good gravity descriptions and
small strips have good brane probe descriptions \cite{Lin:2004nb}. In
the configuration illustrated in figure \ref{figd}, the $n_{B1}$ M2
branes have merged to form a dielectric M5-brane wrapping a $S^3$
section of the A2 4-cycle.

\subsection{Another brane probe domain wall}

Before ending this section, let us also describe another curious brane
configuration in the LLM geometry which also describes a domain wall
interpolating between two distinct vacua.

As a starting point, we recall the fact, originally shown in
\cite{Cheon:2011gv}, that an M2 brane probe localized in the 8
dimensions parametrized by $x$, $y$, $S^3$, and $\tilde S^3$ are
attracted toward $y=0$ with points on $x$ which correspond to the {\it
concave} corners of the Young diagram, i.e.\ where B0 meets A1, B1
meets A2, etc. Anti-M2 branes, on the other hand, are attracted to
$y=0$ with points on $x$ corresponding to the {\it convex} corners. At
these attractor points, the effective tension of the probe branes are
degenerate and saturates the BPS condition. It may seem
counter-intuitive for an anti M2-brane to coexist as a BPS state in
the presence of M2-branes, but such a configuration is made possible
by the presence of fluxes.

The effective potential experienced by the M2 and anti M2-brane probes
is illustrated in figure 5 of \cite{Cheon:2011gv}. In light of the
degenerate minima, say, for the M2 branes corresponding to the convex
corners of the Young diagram, it is natural to contemplate
constructing a kink-like embedding of M2-branes interpolating between
two of these minima.

To set up this exercise, we need to write down the world volume action
for the M2-branes embedded into this background. Alternatively, since
we expect the domain wall to be transitionally invariant in one of the
extended dimensions of $R^{1,2}$, say $z_2$, we can compactify this
dimension and reduce to IIA description where the M2 becomes a
fundamental string. We can then analyze the Nambu-action for the
fundamental string. It would be straightforward to lift the
configuration of fundamental string that we find in type IIA
description to M-theory.

A useful place to read off the IIA background is (D.1) of
\cite{Lin:2004nb}. We will also scale in the parameter $\mu_0$
following \cite{Cheon:2011gv}. If one wishes to consider an embedding
of fundamental string in the $x$ direction as $z_1$ is varied, one can
parametrize the shape of the string by $x(z_1)$. The Nambu action then
takes the form
\be L = \left[ e^{2 \phi} \sqrt{1 + e^{-2 \phi} h^2 \left({d x \over d
z_1}\right)^2} + B\right] \ee
where $h$, $B$, and $\phi$ are functions of $x$. $B$ is an
abbreviation for the $B_{t z_1}$ component of the NSNS 2-form. Now,
since the action does not depend explicitly on $z$, one can infer the
conservation of the Hamiltonian
\be H = {\partial L \over \partial \left({\partial L \over \partial
z_1}\right)} - L = B - {e^{2 \phi}\over \sqrt{1 + e^{-2 \phi} h^2
x'(z_1)^2}} \ . \ee
In order to impose the asymptotic behavior that $x'(z_1) =0$ at $z_1 =
\pm\infty$, we set $H=0$. Then, we find
\be x'(z_1) = \pm {\sqrt{e^{2 \phi} (e^{2 \phi}- B)(e^{2 \phi} +
B)}\over B h} \label{xp}\ee
where right hand side is a function of $x$. One can therefore find
$z_1(x)$ by integrating this equation.

\begin{figure}[t]
\centerline{\includegraphics[width=3in]{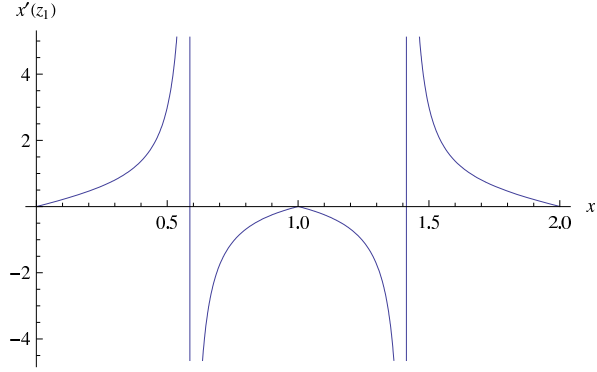}}
\caption{The plot of $x'(z_1)$ as a function of $x$ as found in
(\ref{xp}). \label{fige}}
\end{figure}

When this expression is substituted back into the action, it
simplifies dramatically to
\be T \int dz_1 \, L = T \int dz_1 \, \mu_0 x'(z_1) = T \mu_0 \int dx
, \qquad T = {1 \over 2 \pi l_s^2} \ . \ee
In other words, the tension (which is a mass in the IIA description)
is directly proportional to the extent to which the domain wall
(particle) is extended in the $x$ direction. When applied to the
configuration like the one illustrated in figure \ref{figb}.A, one
finds that $x'(z_1)$ indeed vanishes at all $x$'s at all of the
interfaces of the black and the white regions. In the case where we
set these interfaces at $x=0$, $x=1$, and $x=2$, the $x'(z_1)$ as is
given in (\ref{xp}) as a function of $x$ looks like what is
illustrated in figure \ref{fige}.

\begin{figure}[t]
\centerline{\includegraphics[width=3in]{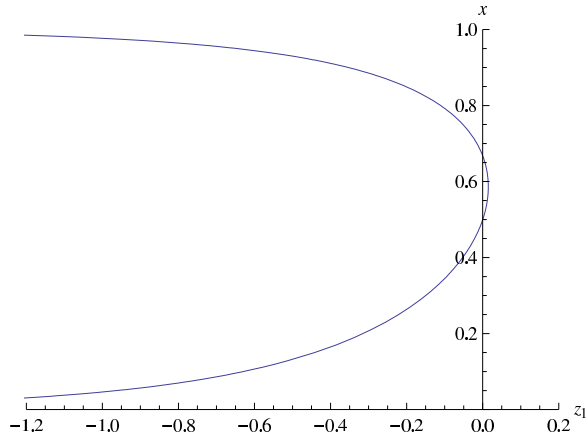}}
\caption{The solution $x(z_1)$ obtained by numerically solving
(\ref{xp}). \label{figf}}
\end{figure}

That $x'(z_1)$ is diverging between the interfaces suggests that the
embedding of $x(z)$ is turning around at that point. Numerically
solving for $x(z_1)$ leads to an embedding illustrated in figure
\ref{figf}.

\begin{figure}[t]
\centerline{\includegraphics{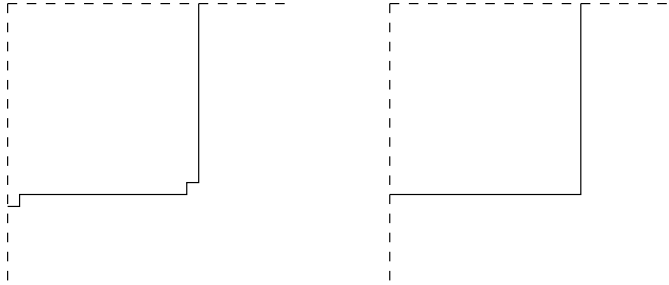}}
\caption{The Young diagram for the vacuum states on two sides of the
domain wall represented by the brane embedding illustrated in figure
\ref{figf}. \label{figg}}
\end{figure}

This picture suggests that in the $z_1 \rightarrow -\infty$ limit, an
M2 and an anti M2-brane probe is embedded on the convex and concave
corners of the Young diagram, as we illustrate on the left side in
figure \ref{figg}. As we increase $z_1$, these M2 and anti M2 brane
merge and disappear, leaving a configuration illustrated by the second
Young diagram in figure \ref{figg}.

The mass $T \mu_0 \Delta x$ which we computed in type IIA lifts
immediately to tension $\tau = T_{M2} \mu_0 \Delta x$ in
M-theory. Since $\Delta x$ is quantized according to (\ref{quant}), we
infer that the tension of the domain wall is
\be \tau = {\mu_0^2 \over 2 \pi} \Delta n \ee
where $\Delta n$ is the quantized length of say the cycle A1 in the
configuration of figure \ref{figg}.

Now, this can be compared against the computation of the domain wall
carried out on the field theory side \cite{Hanaki:2008cu}. From
reading off the field theory vacua from figure \ref{figg}, and
applying the formula (60) in \cite{Hanaki:2008cu}, we read off
\be \tau = {\mu^2 \over 2 \pi} \Delta n \ . \ee
So we learn that the two computations match in the scaling with
respect to $\Delta n$, and are in complete agreement if we identify
$\mu_0$ from the supergravity solution with $\mu$ as defined in
\cite{Gomis:2008vc}. This identification is consistent with what is
reported in \cite{Cheon:2011gv} based on comparing the masses of the
BPS vortecies, except for the fact that \cite{Cheon:2011gv} defines
$\mu$ with an extra power of $2 \pi$ as can be inferred from the form
of the superpotential i.e.\ their (C.18).\footnote{There appears to be
a minor factors of 2 error in (4.7) of \cite{Cheon:2011gv}. The factor
of $(2 \mu_0 / \pi T_{M2})$ should instead read $(\mu_0 / 2 \pi
T_{M2})$. This follows from (2.25) of \cite{Cheon:2011gv}. We thank
Seok Kim for clarification on this point.}

Configuration like the one illustrated in figure \ref{figf} is quite
intriguing. One can introduce a finite mass vortex state connecting
the M2 and the anti M2. Such a state, however, can self-annihilate by
pushing it toward the domain wall. More generally, a configuration
like this appears to offer an interesting laboratory to explore
various aspects of brane/anti-brane physics.

\section{Discussions}

In this article, we considered the effects of the back reaction of
wrapped M5-brane behaving effectively as a domain wall in the
holographic description of certain 2+1 dimensional field
theories. These back reaction effects facilitate the shift in the
vacuum as one crosses from one side of the domain wall to the other in
the gravity description. We studied and confirmed the compatibility
between the vacuum structure implied by the back reaction of the
domain wall brane and the quantization condition on charges and fluxes
intrinsic in these backgrounds. While one expects these issues to
ultimately work out consistently, the fact that it actually does so is
rather non-trivial in light of the fact that quantization condition of
charges and fluxes themselves are quite intricate. Confirming the
consistency between the structures of the quantized fluxes and the
domain walls can therefore be viewed as a diagnostic.

One interesting issue that this analysis clarifies is the fate of the
non-supersymmetric state constructed by starting from the $B_8$ or the
Stenzel background with $Q_2^0=0$ and adding a supersymmetry breaking
M2-brane (or an anti M2-brane).  In the cases where the deep IR is
superconformal, this has the effect of shifting the brane charge
\be Q_2^0 = N - {l(l-k) \over 2k} \ee
to be slightly negative, and one expects the system not flow to an IR
fixed point preserving any supersymmetries. In the case of $B_8$ and
Stenzel, where $Q_2^0$ has an additional term dependent on $q$, one
can change the charges carried by the bulk fields to some number of
supersymmetry preserving M2-branes at the expense of shifting
$q$. Such a shift can happen by crossing the domain wall that we
considered in this article, or though a tunneling process one
constructs by analytically continuing the domain wall solution. Since
these M2-branes can annihilate against the supersymmetry breaking
M2-branes, the system can relax into the supersymmetric vacua.

The only exception to this scenario is the case where the magnitude of
the self-dual 4-form (or the anti-self-dual 4-form) is vanishing.  In
this case, there are no charges carried by the bulk fields that one
can harvest using the domain wall/tunneling transition to annihilate
the supersymmetry breaking M2-brane. Furthermore, if we set the 4-form
and the brane charge to zero, then we are also setting all of the
sources of warping, i.e.\ the Maxwell charge, to zero.  So clearly,
this is a degenerate case.

It is also interesting to see a consistent picture emerging also in
the case of the correspondence between LLM and the mass deformed ABJM
model. What makes this example special is the fact that the field
theory side of the correspondence is better understood than the other
examples.  One curious feature of the LLM geometry on the gravity
side, in contrast to the other examples, is the fact that in the LLM
case, the topology of the 4-cycles are the consequence of the 4-form
fluxes, whereas in the other examples the 4-cycles existed
independently of turning on the flux.  There may be more important
lessons pertaining to dynamical topology changing processes in string
theory related to this observation.

Another interesting feature of these mass deformed theories is that
they appear to support a set of degenerate vacua. It would be useful
to identify a set of physical observables which can be used as a way
to discriminate between these degenerate vacua. On the field theory
side, the expectation value of the superpotential is a partial probe
of this issue. A natural candidate dual of this observable is the
superpotential of Gukov, Vafa, and Witten \cite{Gukov:1999ya}. These
superpotentials should also be useful for computing the tension of
domain walls. It would be interesting to demonstrate a more detailed
correspondence between the superpotential of the field theory side and
the GVW superpotential, especially for the case of LLM/mass deformed
ABJM theory where a lot is known already on the field theory side.

\section*{Acknowledgements}

We would like to thank Ofer Aharony, Shinji Hirano, and Peter Ouyang
for collaboration on related issues and for discussions at the early
stage of this work and Horatiu Nastase for a discussion on fuzzy 3
spheres and mass deformed ABJM theories.  We also thank
Seok Kim, Igor Klebanov, Oleg Lunin, Peter Ouyang, and Diego Trancanelli
for helpful comments and discussions. This work was supported in part
by the DOE grant DE-FG02-95ER40896.

\bibliography{domainwall}\bibliographystyle{utphys}

\providecommand{\href}[2]{#2}\begingroup\raggedright\begin{thebibliography}{10}

\bibitem{Aharony:2008ug}
O.~Aharony, O.~Bergman, D.~L. Jafferis, and J.~Maldacena, ``{${\cal N}=6$
  superconformal Chern-Simons-matter theories, M2-branes and their gravity
  duals},'' {\em JHEP} {\bf 0810} (2008) 091,
  \href{http://www.arXiv.org/abs/0806.1218}{{\tt 0806.1218}}.

\bibitem{Marolf:2000cb}
D.~Marolf, ``{Chern-Simons terms and the three notions of charge},''
\href{http://www.arXiv.org/abs/hep-th/0006117}{{\tt hep-th/0006117}}.
%%CITATION = HEP-TH/0006117;%%.

\bibitem{Aharony:2009fc}
O.~Aharony, A.~Hashimoto, S.~Hirano, and P.~Ouyang, ``{D-brane charges in
  gravitational duals of 2+1 dimensional gauge theories and duality
  cascades},'' {\em JHEP} {\bf 01} (2010) 072,
\href{http://www.arXiv.org/abs/0906.2390}{{\tt 0906.2390}}.
%%CITATION = 0906.2390;%%.

\bibitem{Hashimoto:2008iv}
A.~Hashimoto and P.~Ouyang, ``{Supergravity dual of Chern-Simons Yang-Mills
  theory with ${\cal N}=6,8$ superconformal IR fixed point},'' {\em JHEP} {\bf
  10} (2008) 057,
\href{http://www.arXiv.org/abs/0807.1500}{{\tt 0807.1500}}.
%%CITATION = 0807.1500;%%.

\bibitem{Hashimoto:2010bq}
A.~Hashimoto, S.~Hirano, and P.~Ouyang, ``{Branes and fluxes in special
  holonomy manifolds and cascading field theories},''
\href{http://www.arXiv.org/abs/1004.0903}{{\tt 1004.0903}}.
%%CITATION = 1004.0903;%%.

\bibitem{Bena:2010gs}
I.~Bena, G.~Giecold, and N.~Halmagyi, ``{The backreaction of anti-M2 branes on
  a warped Stenzel space},'' {\em JHEP} {\bf 04} (2011) 120,
\href{http://www.arXiv.org/abs/1011.2195}{{\tt 1011.2195}}.
%%CITATION = 1011.2195;%%.

\bibitem{Gukov:2001hf}
S.~Gukov and J.~Sparks, ``{M-theory on $Spin(7)$ manifolds. I},'' {\em Nucl.
  Phys.} {\bf B625} (2002) 3--69,
\href{http://www.arXiv.org/abs/hep-th/0109025}{{\tt hep-th/0109025}}.
%%CITATION = HEP-TH/0109025;%%.

\bibitem{Martelli:2009ga}
D.~Martelli and J.~Sparks, ``{$AdS_4/CFT_3$ duals from M2-branes at
  hypersurface singularities and their deformations},'' {\em JHEP} {\bf 12}
  (2009) 017,
\href{http://www.arXiv.org/abs/0909.2036}{{\tt 0909.2036}}.
%%CITATION = 0909.2036;%%.

\bibitem{Lin:2004nb}
H.~Lin, O.~Lunin, and J.~M. Maldacena, ``{Bubbling AdS space and 1/2 BPS
  geometries},'' {\em JHEP} {\bf 0410} (2004) 025,
  \href{http://www.arXiv.org/abs/hep-th/0409174}{{\tt hep-th/0409174}}.

\bibitem{Gomis:2008vc}
J.~Gomis, D.~Rodriguez-Gomez, M.~Van~Raamsdonk, and H.~Verlinde, ``{A massive
  study of M2-brane proposals},'' {\em JHEP} {\bf 0809} (2008) 113,
  \href{http://www.arXiv.org/abs/0807.1074}{{\tt 0807.1074}}.

\bibitem{Gukov:1999ya}
S.~Gukov, C.~Vafa, and E.~Witten, ``{CFT's from Calabi-Yau four-folds},'' {\em
  Nucl. Phys.} {\bf B584} (2000) 69--108,
\href{http://www.arXiv.org/abs/hep-th/9906070}{{\tt hep-th/9906070}}.
%%CITATION = HEP-TH/9906070;%%.

\bibitem{Hashimoto:2011aj}
A.~Hashimoto and P.~Ouyang, ``{Quantization of charges and fluxes in warped
  Stenzel geometry},''
\href{http://www.arXiv.org/abs/1104.3517}{{\tt 1104.3517}}.
%%CITATION = 1104.3517;%%.

\bibitem{Freed:1999vc}
D.~S. Freed and E.~Witten, ``{Anomalies in string theory with D-branes},''
\href{http://www.arXiv.org/abs/hep-th/9907189}{{\tt hep-th/9907189}}.
%%CITATION = HEP-TH/9907189;%%.

\bibitem{Witten:1996md}
E.~Witten, ``{On flux quantization in M-theory and the effective action},''
  {\em J. Geom. Phys.} {\bf 22} (1997) 1--13,
\href{http://www.arXiv.org/abs/hep-th/9609122}{{\tt hep-th/9609122}}.
%%CITATION = HEP-TH/9609122;%%.

\bibitem{Kachru:2002gs}
S.~Kachru, J.~Pearson, and H.~L. Verlinde, ``{Brane/Flux annihilation and the
  string dual of a non- supersymmetric field theory},'' {\em JHEP} {\bf 06}
  (2002) 021,
\href{http://www.arXiv.org/abs/hep-th/0112197}{{\tt hep-th/0112197}}.
%%CITATION = HEP-TH/0112197;%%.

\bibitem{Klebanov:2010qs}
I.~R. Klebanov and S.~S. Pufu, ``{M-Branes and metastable states},''
\href{http://www.arXiv.org/abs/1006.3587}{{\tt 1006.3587}}.
%%CITATION = 1006.3587;%%.

\bibitem{Gibbons:1989er}
G.~W. Gibbons, D.~N. Page, and C.~N. Pope, ``{Einstein metrics on $S^3$, $R^3$
  and $R^4$ bundles},'' {\em Commun. Math. Phys.} {\bf 127} (1990)
529.
%%CITATION = CMPHA,127,529;%%.

\bibitem{Bryant:1989mv}
R.~Bryant and S.~Salamon, ``{On the construction of some complete metrices with
  expectional holonomy},'' {\em Duke Math. J.} {\bf 58} (1989)
829.
%%CITATION = DUMJA,58,829;%%.

\bibitem{Cvetic:2001pga}
M.~Cvetic, G.~W. Gibbons, H.~Lu, and C.~N. Pope, ``{New complete non-compact
  $Spin(7)$ manifolds},'' {\em Nucl. Phys.} {\bf B620} (2002) 29--54,
\href{http://www.arXiv.org/abs/hep-th/0103155}{{\tt hep-th/0103155}}.
%%CITATION = HEP-TH/0103155;%%.

\bibitem{Cvetic:2001zx}
M.~Cvetic, G.~W. Gibbons, H.~Lu, and C.~N. Pope, ``{Cohomogeneity one manifolds
  of $Spin(7)$ and $G_2$ holonomy},'' {\em Phys. Rev.} {\bf D65} (2002) 106004,
\href{http://www.arXiv.org/abs/hep-th/0108245}{{\tt hep-th/0108245}}.
%%CITATION = HEP-TH/0108245;%%.

\bibitem{Duff:1983nu}
M.~J. Duff, B.~E.~W. Nilsson, and C.~N. Pope, ``{Spontaneous supersymmetry
  breaking by the squashed seven sphere},'' {\em Phys. Rev. Lett.} {\bf 50}
  (1983)
2043.
%%CITATION = PRLTA,50,2043;%%.

\bibitem{Duff:1986hr}
M.~J. Duff, B.~E.~W. Nilsson, and C.~N. Pope, ``{Kaluza-Klein supergravity},''
  {\em Phys. Rept.} {\bf 130} (1986)
1--142.
%%CITATION = PRPLC,130,1;%%.

\bibitem{Witten:1998xy}
E.~Witten, ``{Baryons and branes in anti de Sitter space},'' {\em JHEP} {\bf
  07} (1998) 006,
\href{http://www.arXiv.org/abs/hep-th/9805112}{{\tt hep-th/9805112}}.
%%CITATION = HEP-TH/9805112;%%.

\bibitem{Stenzel}
M.~Stenzel, ``{Ricci-flat metrics on the complexification of a compact ranke
  one symmetric space},'' {\em Manuscr. Math.} {\bf 80} (1993)
151--163.
%%CITATION = MSMHB,80,151;%%.

\bibitem{Cvetic:2000db}
M.~Cvetic, G.~W. Gibbons, H.~Lu, and C.~N. Pope, ``{Ricci-flat metrics,
  harmonic forms and brane resolutions},'' {\em Commun. Math. Phys.} {\bf 232}
  (2003) 457--500,
\href{http://www.arXiv.org/abs/hep-th/0012011}{{\tt hep-th/0012011}}.
%%CITATION = HEP-TH/0012011;%%.

\bibitem{Cheon:2011gv}
S.~Cheon, H.-C. Kim, and S.~Kim, ``{Holography of mass-deformed M2-branes},''
  \href{http://www.arXiv.org/abs/1101.1101}{{\tt 1101.1101}}.

\bibitem{Hosomichi:2008jb}
K.~Hosomichi, K.-M. Lee, S.~Lee, S.~Lee, and J.~Park, ``{${\cal N}=5,6$
  superconformal Chern-Simons theories and M2-branes on Orbifolds},'' {\em
  JHEP} {\bf 0809} (2008) 002, \href{http://www.arXiv.org/abs/0806.4977}{{\tt
  0806.4977}}.

\bibitem{Kim:2010mr}
H.-C. Kim and S.~Kim, ``{Supersymmetric vacua of mass-deformed M2-brane
  theory},'' {\em Nucl.Phys.} {\bf B839} (2010) 96--111,
  \href{http://www.arXiv.org/abs/1001.3153}{{\tt 1001.3153}}.

\bibitem{Hanaki:2008cu}
K.~Hanaki and H.~Lin, ``{M2-M5 systems in ${\cal N}=6$ Chern-Simons theory},''
  {\em JHEP} {\bf 0809} (2008) 067,
  \href{http://www.arXiv.org/abs/0807.2074}{{\tt 0807.2074}}.

\end{thebibliography}\endgroup

\end{document}